\documentclass[aps,prl,reprint,groupedaddress,preprintnumbers]{revtex4-1}
\usepackage{amsmath,amssymb,mathtools}
\usepackage{graphicx}
\usepackage{hyperref}
\usepackage{subfigure}
\usepackage{comment}
\usepackage{tensor}
\usepackage{cases}
\usepackage{color}
\usepackage{soul}
\usepackage{bm}

\newcommand{\I}{{\rm s}}
\newcommand{\II}{{\rm d}}
\newcommand{\F}{{\rm F}}
\newcommand{\R}{{\rm R}}
\newcommand{\be}{\begin{eqnarray}}
\newcommand{\ee}{\end{eqnarray}}
\newcommand{\n}{\nonumber \\}
\begin{document}

\preprint{KEK-TH-1970, HUPD-1703}
\vspace{2cm}
\title{Entanglement-induced quantum radiation}
\vspace{1cm}

\author{Satoshi Iso}
\affiliation{KEK Theory Center, High Energy Accelerator Research Organization (KEK), 
Tsukuba, Ibaraki 305-0801, Japan}

\author{Rumi Tatsukawa}
\affiliation{Graduate school of Physical Sciences, Department of Physical Sciences,
Hiroshima University, Higashi-hiroshima, Kagamiyama 1-3-1, 739-8526, Japan}

\author{Kazushige Ueda}
\affiliation{Graduate school of Physical Sciences, Department of Physical Sciences,
Hiroshima University, Higashi-hiroshima, Kagamiyama 1-3-1, 739-8526, Japan}

\author{Kazuhiro Yamamoto}
\affiliation{Graduate school of Physical Sciences, Department of Physical Sciences,
Hiroshima University, Higashi-hiroshima, Kagamiyama 1-3-1, 739-8526, Japan}

\begin{abstract} 
Quantum entanglement of the Minkowski vacuum state 
between left and right Rindler wedges generates thermal behavior 
in the right Rindler wedge, which
is known as the Unruh effect. In this letter, we show that there is another consequence of this entanglement, namely
entanglement-induced quantum radiation emanating from a uniformly accelerated object. 
We clarify why it is in agreement with our intuition that 
 incoming and outgoing energy fluxes should cancel each other out in a thermalized state.

\end{abstract} 

\maketitle
{\it Introduction.}--- A major outstanding question in quantum field theory is whether an Unruh-de 
Witt detector (which is a uniformly accelerated object coupled to a radiation field)
 would emit radiation. 
 A detection of such radiation, if it exists, would have a huge impact upon fundamental physics, and
various experimental proposals toward  detecting the radiation  have been made 
\cite{ChenTajima,Schutzhold,Schutzhold2,ELI,LH,IYZ,Lin}. 
After an infinitely long time, the Unruh-de Witt detector would become thermalized to 
the Unruh temperature \cite{Unruh}, and, in analogy with an ordinary equilibrium system in a thermal bath,
one may intuitively expect that the incoming and outgoing energy fluxes
from the accelerated object should cancel out, resulting in a net flux of zero. 

This issue has been debated for a long time.
In a toy model with $d=1+1$ dimensions, \cite{HuRaval} showed that there is no net flux,
which is consistent with our intuition.
For a charged particle model in $d=3+1$ dimensions, Refs.~\cite{IYZ,OYZ15,OYZ16} have shown that, although most of the terms cancel out
when interference effects are carefully considered, 
there remains some radiation.
However, these calculations could be performed only approximately, leaving it unclear whether this radiation exists.
The authors of \cite{LH, IOTYZ} showed that, in a toy model in $d=3+1$ dimensions where we can solve the equation of motion exactly, 
the cancellation is not exact and quantum radiation emanates; thus, there must be an error in our intuition.

\begin{figure}[t]
\begin{center}
    \includegraphics[width=6cm]{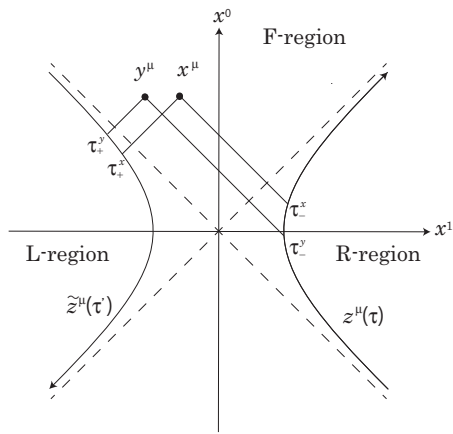}
    \caption{$\tau_-$ is the proper time 
on the trajectory $x=z(\tau)$, whereas $\tau_+$ is that on the
    imaginary trajectory in the L-region.
\label{fig:tau}
}
\end{center}
\end{figure}

In our previous paper \cite{IOTYZ}, we presented some circumstantial evidence that this radiation is responsible 
for a nonlocal correlation of the quantum field in the Minkowski vacuum state, which originates 
in the entanglement between the
left (L) and right (R) Rindler wedges. In this letter, we explicitly show that this is indeed the case
and explain why it does not contradict to our intuition that net radiation should cancel out 
in a thermalized system.
More specifically, the origin of the radiation is entanglement between the wave functions (Rindler modes) in the R-region
and the right-moving waves (Kasner modes) in the F-region in Fig \ref{fig:tau}.

\par
{\it Radiation from an Unruh-de Witt detector.}--- We consider a toy model in $d=3+1$ dimensions, a coupled system of 
a harmonic oscillator, $Q$, accelerating uniformly and a massless scalar field, $\phi(x)$.
The action is given by
\begin{eqnarray}  
S[Q,\phi ; z] &=&
\frac{m}{2} \int d \tau \bigl( \dot{Q}^2(\tau) - \Omega_0^2 Q^2(\tau) \bigr) 
\nonumber\\
&+& {1\over 2}\int d^4 x \partial^\mu \phi(x) \partial_\mu \phi(x)  
\nonumber\\
&+& \lambda \int d^4 x d\tau Q(\tau) \phi(x) \delta^{(4)}_D(x-z(\tau)),
\end{eqnarray} 
with the world-line trajectory $z^\mu(\tau)$ in the R-region (see Fig. \ref{fig:tau}) at a uniform acceleration $a$,
\be
z^\mu(\tau)=a^{-1}\left(\sinh{a\tau},\cosh{a\tau},0,0\right) . 
\ee
The equations of motion can be solved exactly \cite{LH,IYZ}, and
after an infinitely long time, the system relaxes to an equilibrium state.

In the presence of the harmonic oscillator (which is our Unruh-de Witt detector) $Q(\tau)$, the field $\phi(x)$ in the Heisenberg picture
is given by a sum of homogeneous  $\phi_{\rm h}(x)$ and inhomogeneous  $\phi_{\rm inh}(x)$ terms,
 \be
  \phi(x)=\phi_{\rm h}(x)+\phi_{\rm inh}(x) .
  \ee
The homogeneous term $\phi_{\rm h}(x)$ is nothing but the vacuum fluctuation of the field at the position $x$,
while $\phi_{\rm inh}(x)$ can be 
solved in terms of the $Q$ operator as
\begin{eqnarray}
\phi_{\rm inh}(x) =\lambda \int d\tau Q(\tau)G_\R(x-z(\tau)),
\label{phiQG}
\end{eqnarray}
where $G_R(x-y)$ is the retarded Green's function of the d'Alembertian operator.
$\phi_{\rm inh}(x)$ can be further written in terms 
of the field $\phi_h(z(\tau))$ on the trajectory  $x=z(\tau)$ in the R-region,
by using the solution of the equation of motion for $Q(\tau)$
\begin{eqnarray}
\left( \partial_\tau^2 + 2 \gamma \partial_\tau + \Omega^2 \right) Q(\tau) = (\lambda/m) \phi_{\rm h}(z(\tau)) .
\label{eqofQ}
\end{eqnarray}
Here $\gamma=\lambda^2/8\pi m$ and $\Omega$ is the renormalized frequency 
(see Ref.~\cite{LH,IYZ}).

Thus, when we calculate the energy flux of radiation in the F-region, namely 
 the two-point correlation function of fields $\langle \phi(x) \phi(y)\rangle$  at $x,y \in \text{F-region}$;
it is  necessary to calculate the two-point correlation functions of the field $\phi_{\rm h}(x)$ in the F-region
and the field $\phi_{\rm h}(y)$ on the trajectory $y=z(\tau)$ in the R-region (or both in the R-region).

Let us look at this more explicitly. 
The quantity we are interested in is the two-point correlation function 
$ \langle\phi_{\rm }(x)\phi_{\rm }(y)\rangle$
in the F-region ($x, y \in \text{F}$), since
the future light-cone of the detector is almost in the F-region in the 
the large distance limit and non-vanishing radiation flux is expected to appear there.
By subtracting the vacuum correlation function $\langle\phi_{\rm h}(x)\phi_{\rm h}(y)\rangle$, 
it is given by the sum of three terms,
\begin{eqnarray}
&& \langle\phi_{\rm }(x)\phi_{\rm }(y)\rangle-\langle\phi_{\rm h}(x)\phi_{\rm h}(y)\rangle =
\langle\phi_{\rm inh}(x)\phi_{\rm inh}(y)\rangle
\n
&&~~+ \langle\phi_{\rm h}(x)\phi_{\rm inh}(y)\rangle
+\langle\phi_{\rm inh}(x)\phi_{\rm h}(y)\rangle  .
\label{pppppp}
\end{eqnarray}
We call the first term on the right-hand side a {\it naive radiation} term;
its detection was the target of the original experimental proposals.
The second and the third terms relate to the {\it interference}.
The existence or non-existence of the Unruh radiation is determined by 
how much of the naive radiation 
is canceled by the interference terms (Refs.~\cite{IYZ,IOTYZ}).

As we noted, $\phi_{\rm inh}(x)$ is determined by the vacuum fluctuation $\phi_{\rm h}(z(\tau))$ on the trajectory in the R-region.
Thus, the naive radiation term is calculated in terms of the correlation function of $\phi_{\rm h}(z(\tau))$ and $\phi_{\rm h}(z(\tau'))$,
both of which are in the R-region.
On the other hand, the interference terms are
calculated in terms of
$\langle\phi_{\rm h}(x)\phi_{\rm h}(z(\tau))\rangle$, where $x \in {\text F}$ and 
 $z(\tau)$ is the trajectory of the detector in the R-region.
 Thus, the interference terms reflect the correlations between the vacuum fluctuations in the F- and R-regions.
A straightforward calculation shows that \cite{IYZ,IOTYZ}
\begin{eqnarray}
  &&
  \hspace{-5mm}
  \langle\phi_{\rm h}(x)\phi_{\rm h}(z(\tau))\rangle=
-{i\over 8\pi^2\rho_0(x)}\int_{-\infty}^{+\infty}d\omega e^{-i\omega\tau}
\nonumber\\
&&~~~~~~\Bigl(
{e^{\pi\omega/a}\over e^{2\pi\omega/a}-1}e^{i\omega\tau_+^x}
-{1\over e^{2\pi\omega/a}-1}e^{i\omega\tau_-^x}
\Bigr),
\label{resultoffs}
\end{eqnarray}
where $\rho_0(x)$ is the distance between the position $x$ and the detector defined in Eq.(3.1) in \cite{IOTYZ},
and $\tau_\pm^x$ are the proper times depicted in Fig. \ref{fig:tau}. 

The second term in parentheses in Eq.(\ref{resultoffs}), which depends on
 $\tau_-^x$ on the real trajectory, is shown to completely cancel the naive radiation term \cite{IOTYZ}.
Thus, our intuition concerning the absence of naive radiation in a thermalized system seems to be correct so far.
However, the interference terms have an additional contribution, namely 
 the first term in parentheses in Eq.(\ref{resultoffs}).
We showed that this term generates a new type of quantum radiation from the accelerated object in the R-region. 
 This term differs from the second one in two ways. First, it is a function of $\tau_+^x$, which is a 
 proper time on the imaginary trajectory in the L-region. Second, 
the first term in parentheses in Eq.(\ref{resultoffs})
is multiplied by an additional factor, $e^{\pi \omega/a}$. In \cite{IOTYZ}, we discussed how these two points indicate that
  a new type of radiation is induced by entanglement of the vacuum state.
 We call it {\it entanglement-induced quantum radiation}.
 In the rest of this letter, we make this radiation more transparent by scrutinizing the Hilbert 
space structure in the F-region. \\

\begin{figure}[t]
\begin{center}
    \includegraphics[width=6cm]{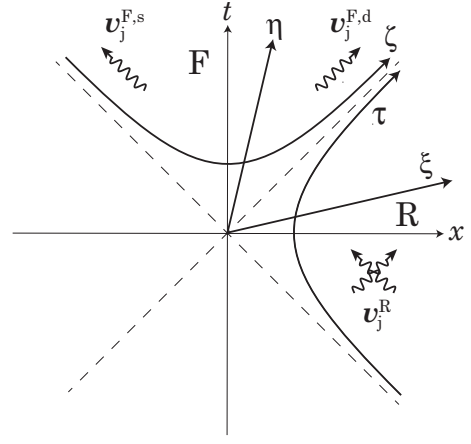}
    \caption{Coordinates of the right Rindler wedge (R-region, $x>|t|$) and the 
expanding degenerate Kasner universe (F-region, $t>|x|$). Here, $v_j^{\rm R}, v_j^{\rm F,s},$ and $v_j^{\rm F,\rm d}$
indicate the Rindler mode, the left-moving-wave Kasner mode, and the right-moving-wave Kasner mode, 
respectively. The Rindler mode is a standing-wave mode in the R-region.
\label{fig:coordinate}}
\end{center}
\end{figure}

{\it  Minkowski vacuum as an entangled state.}---
The Minkowski vacuum state of $\phi(x)$ can be written as an
entangled state in terms of the L and R Rindler states, as shown by \cite{UnruhWald}
\begin{eqnarray}
|0,{\rm M}\rangle
=\prod_j\biggl[N_j\sum_{n_j=0}^\infty e^{-\pi n_j \omega_j/a}
|n_j \rangle_{{\rm R}}
\otimes|n_j \rangle_{\rm L} \biggr]  ,
\label{Minkowskivac}
\end{eqnarray}
with $j=(\omega,\bm k_\perp)$ and $N_j=\sqrt{1-e^{-2\pi \omega/a}}$.
The quantized field in the R-region is expanded \cite{Higuchi} as
\begin{eqnarray}
  \phi_{\rm h} (x)=\int_{0}^{\infty} d \omega \int d^2 k_\perp \left( {\hat a}_{\omega,{\bm k}_\perp}^{\rm R}
  v_{\omega,{\bm k}_\perp}^{\rm R}(x)
+{\rm h.c.}
\right) ,
\label{qfR}
\end{eqnarray}
where $x \in \text{R-region}$ and
\begin{eqnarray}
v_{\omega,{\bm k}_\perp}^\R(x)=\sqrt{\sinh \pi \omega/a\over 4\pi^4 a}
K_{i\omega/a}\Bigl({\kappa e^{a\xi}\over a}\Bigr)e^{i\bm k_\perp\cdot \bm x_\perp-i\omega\tau}.
\label{mfR}
\end{eqnarray}
Here,  $\kappa=|\bm k_\perp|$,  and the 
Rindler coordinates $(\tau,\xi)$ are defined by
$t={a^{-1}}e^{a\xi}\sinh a\tau$,~~$x={a^{-1}}e^{a\xi}\cosh a\tau$.
The line element of a flat Minkowski spacetime is given by
$ ds^2=e^{2a\xi}(d\tau^2-d\xi^2)-d\bm x_\perp^2$ in Rindler coordinates.
The coordinates cover the R-region, a quarter of the Minkowski spacetime in Fig.~\ref{fig:tau} and Fig.~\ref{fig:coordinate}.
The state $|n_j \rangle_\text{R}$ is an $n_j$-th excited state created by the operator $ {\hat a}_{\omega,{\bm k}_\perp}^{\rm R \dagger}$. 
The state  $|n_j \rangle_\text{L}$ is similarly defined in the L-region with the sign of its momentum reversed for convenience.
Then, as in (\ref{Minkowskivac}),
the Minkowski vacuum state is written as the entangled state of the
excited states in the L and R regions.

The Minkowski vacuum can be similarly written in the F-region.
The F-region ($t>|x|$) is described by the expanding degenerate Kasner universe
with coordinates $(\eta,~\zeta)$. 
$x$ and $t$ are respectively given by $t={a^{-1}}e^{a\eta}\cosh a\zeta$ and~~$x={a^{-1}}e^{a\eta}\sinh a\zeta$,
and the line element is
$ds^2=e^{2a\eta}(d\eta^2-d\zeta^2)-d\bm x_\perp^2$. 
In the F-region, the quantized field $\phi_{\rm h}(x)$
can be expanded \cite{Higuchi} as
\begin{eqnarray}
&&\hspace{-10mm}
\phi_{\rm h} (x)=\int_{-\infty}^{+\infty} 
d \omega\int  d^2k_\perp \left( {\hat a}_{\omega,{\bm k}_\perp}^\F v_{\omega,{\bm k}_\perp}^\F(x)
+{\rm h.c.}
\right), 
\label{qfF}
\end{eqnarray}
where $x \in \text{F}$ and 
\begin{eqnarray}
v_{\omega,{\bm k}_\perp}^\F(x)={-ie^{i\omega\zeta}\over 2\pi\sqrt{4a\sinh (\pi |\omega|/a)}}
J_{-i|\omega|/a}\Bigl({\kappa e^{a\eta}\over a}\Bigr)
e^{i\bm k_\perp\cdot \bm x_\perp}  .
\nonumber\\
\label{mfF}
\end{eqnarray}
It is important to note that 
$\omega$ corresponds to the momentum in the $\zeta$ direction
and takes either a positive or a negative value.
A positive $\omega$ represents a right-moving Kasner mode, whereas a negative
$\omega$ represents a left-moving Kasner mode. 
We then separate the field $\phi_{\rm h}(x)$ in the F-region into  $\phi_{\rm F}^{\rm d}(x)$ with
$\omega>0$ (right-moving) modes and $\phi_{\rm F}^{\rm s}(x) $ with $\omega<0$ (left-moving) 
modes as
\begin{eqnarray}
  &&\phi_{\rm h} (x)=\phi_\F^\II(x)+\phi_\F^\I(x),
\label{phiFF}
\end{eqnarray}
where $x \in \text{F}$. Here, the indices "d" and "s"
indicate {\it dexter} (right) and {\it sinister} (left) in Latin, respectively. 
The field $\phi_{\rm F}^{\rm d}$ is defined by $\phi_{\rm h}(x)$, whose $\omega$ is restricted to $\omega>0$:
\be
\phi_\F^\II(x)= \int_{0}^{+\infty} 
d \omega\int  d^2 k_\perp \Bigl( {\hat a}_{\omega,{\bm k}_\perp}^{\rm F, d} v_{\omega,{\bm k}_\perp}^{\F,\rm d}(x)
+{\rm h.c.}
\Bigr), 
\nonumber
\ee
where $ v_{\omega,{\bm k}_\perp}^{\F,\II}(x)=v_{\omega, -{\bm k}_\perp}^\F(x)$ and 
$ \hat a^{\rm F,\rm d}_{\omega,\bm k_\perp}=\hat a^\F_{\omega,-\bm k_\perp}.$
The transverse momenta are reversed for later convenience.
Another field, $\phi_{\rm F}^{\rm s}(x)$, is defined as
\be
\phi_\F^\I(x)= \int_{0}^{+\infty} 
 d \omega\int  d^2 k_\perp \Bigl({\hat a}_{\omega,{\bm k}_\perp}^{\rm F,s}v_{\omega,{\bm k}_\perp}^{\F,\rm s}(x)
+ {\rm h.c.}
\Bigr),
\nonumber
\ee
where 
$v_{\omega,{\bm k}_\perp}^{\F,\I}(x)=v_{-\omega, {\bm k}_\perp}^\F(x)$
and $\hat a^{\rm F, s}_{\omega,\bm k_\perp}=\hat a^\F_{-\omega,  \bm k_\perp}.$

The vacuum state with respect to these annihilation operators
is defined by $|0,\I\rangle_{\rm F} \otimes|0,\II\rangle_{\rm F}$, and their $n_j$-th excited states
are represented
 by $ |n_j,{\I} \rangle_{\rm F}$ and $|n_j,{\II}\rangle_{\rm F}$. 
In terms of these Kasner modes, the Minkowski vacuum can be written as
\begin{eqnarray}
\hspace{-5mm}
  |0,{\rm M}\rangle=\prod_j\Bigl[
N_j\sum_{n_j=0}^\infty e^{-\pi n_j \omega/a }|n_j,{\I}\rangle_{\rm F} \otimes|n_j,{\II}\rangle_{\rm F}
\Bigr]  ,
\label{MIII}
\end{eqnarray}
which expresses the cosmological quantum entanglement in the F-region \cite{EMM}. 
Notice that in this expression,  $\omega$ in $j=(\omega,\bm k_\perp)$ is restricted to  positive values.

In order to understand the physical origin of entanglement-induced quantum 
radiation given by the first term in parentheses in Eq.(\ref{resultoffs}),
we need to know how the quantum fields $\phi_{\rm h}(x)$  of the  F and R regions are correlated.
 Namely, we need to relate two expressions of the Minkowski vacuum:
  Eq.(\ref{Minkowskivac}) in the
 R-region and Eq.(\ref{MIII}) in the F-region.
For this purpose, we use the following properties that relate the wave functions of 
the Rindler and Kasner modes.
As discussed in Ref.~\cite{Higuchi}, 
the Rindler-mode function in the R-region, $v_{\omega,{\bm k}_\perp}^{\rm R}(x)$, 
can be identified with the left-moving mode $v_{\omega,\bm k_\perp}^{\rm F,s}(x)$ 
in the F-region by a continuation through the Minkowski-mode functions, which properly
describe the evolution across the horizons.
Similarly, the Rindler-mode function in the L-region, 
$v_{\omega,{\bm k}_\perp}^{\rm L}(x)$, is continued to the right-moving 
mode $v_{\omega,\bm k_\perp}^{\rm F,\rm d}(x)$ in the F-region. 
The observation of continuations of the wave functions suggest that a state created by
the operator $( {\hat a}_{\omega,{\bm k}_\perp}^{\rm R})^\dagger$ in the R-region propagates into the F-region
and becomes identified with a state created by the operator 
$(\hat{a}^{\rm F, s}_{\omega,\bm k_\perp})^\dagger$ in the F-region.
Intuitively, this is quite natural. Further detailed analysis is under investigations.  \\

{\it Entanglement-induced quantum radiation.}---
We now calculate the two-point correlation function (\ref{resultoffs}) of fields in the R and F regions
using the operator formalism based on the Hilbert-space structure discussed above.
Since $x \in \text{F}$, we decompose $\langle\phi_{\rm h}(x)\phi_{\rm h}(z(\tau))\rangle$ into the terms
 $\langle \phi_\F^\I(x)\phi_\R(z(\tau)) \rangle$ and $\langle \phi_\F^{\rm d}(x)\phi_\R(z(\tau)) \rangle$ using  (\ref{phiFF}).

First, we evaluate $\langle \phi_\F^\I(x)\phi_\R(z(\tau)) \rangle$. The field $ \phi_\F^\I(x)$ contains
the left-moving modes in the F-region, which are identified with the modes in the R-region as discussed above.
Thus, we can obtain the relations
\be
\langle0,{\rm M}|\hat a_{\omega,{\bm k}_\perp}^{\rm F,s \dagger}\hat a_{\omega',{\bm k}_\perp'}^{\rm R}|0,{\rm M}\rangle
= {\delta(\omega-\omega')\delta^2(\bm k_\perp-\bm k_\perp')\over e^{2\pi\omega/a}-1}, 
\nonumber
\ee
and
\be
 \langle0,{\rm M}|\hat a_{\omega,{\bm k}_\perp}^{\rm F,s}\hat a_{\omega',{\bm k}_\perp'}^{\rm R \dagger}|0,{\rm M}\rangle
={\delta(\omega-\omega')\delta^2(\bm k_\perp-\bm k_\perp') \over 1- e^{-2\pi\omega/a} },  \nonumber
\ee
which reflect the fact that the left-moving modes in the F-region are in thermal equilibrium at temperature $T=a/2\pi$.
Then, $\langle \phi_\F^\I(x)\phi_\R(z(\tau)) \rangle$ can be evaluated as
\begin{eqnarray}
  &&\hspace{-3mm}
\langle0,{\rm M}|\phi_\F^\I(x)\phi_\R(z(\tau))|0,{\rm M}\rangle
  =\int_0^\infty d\omega\int d^2k_\perp
    \nonumber\\
  &&
   \Bigl( {v^{\F,\I*}_{\omega\bm k_\perp}(x)v^{\R}_{\omega\bm k_\perp}(z(\tau))\over e^{2\pi\omega/a}-1}
    +    {v^{\F,\I}_{\omega\bm k_\perp}(x)v^{\R*}_{\omega\bm k_\perp}(z(\tau))\over 1-e^{-2\pi\omega/a}}  \Bigr) .
\label{vvcc}
\end{eqnarray}
The integrations can be performed using the relation
$\int_0^{2\pi} d\varphi e^{i\kappa x_\perp\cos\varphi}=2\pi J_0(\kappa x_\perp)$ and the
 mathematical formulae in \cite{Tables, Magnus} as follows:
\begin{eqnarray}
&&\hspace{-2mm}\int_0^\infty d\kappa\kappa J_{-i\omega/a}\bigl({\kappa\over a}e^{a\eta}\bigr)K_{-i\omega/a}\bigl({\kappa\over a}\bigr)
  J_0(\kappa x_\perp)={ae^{i\omega\tau_+^x-i\omega\zeta}\over 2\rho_0(x)}.
\nonumber
\end{eqnarray}
Then, noting the relation $e^{i\omega\tau_+^{x}-i\omega\zeta} =e^{-i\omega\tau_-^{x}+i\omega\zeta}$,
we can show that $\langle \phi_\F^\I(x)\phi_\R(z(\tau)) \rangle$ is reduced to the 
second term in parentheses in Eq.~(\ref{resultoffs}), canceling out the naive radiation produced by 
the correlation function $\langle\phi_{\rm inh}(x)\phi_{\rm inh}(y)\rangle$.
This result is consistent with our intuition that the net radiation should cancel out in an equilibrium state in
a thermal bath. The naive radiation is completely canceled out by the interference term of the left-moving modes in the F-region.
The modes come from the R-region, and the cancellation shows that if we restrict the mode functions 
of the field $\phi(x)$ in the F-region to those coming from the R-region, no net radiation will emanate from an accelerating object. 

Note, however, that there is an additional contribution.
We next evaluate $\langle \phi_\F^{\rm d}(x)\phi_\R(z(\tau)) \rangle$;
this provides an additional contribution to the correlation function in the F-region. The calculations can be similarly
performed. In this case, we use the relations
\begin{eqnarray}
&& \langle0,{\rm M}|\hat a_{\omega,{\bm k}_\perp}^{\rm F,\rm d}\hat a_{\omega',{\bm k}_\perp'}^{\rm R}|0,{\rm M}\rangle
=\langle0,{\rm M}|\hat a_{\omega,{\bm k}_\perp}^{\rm F,\rm d \dagger}\hat a_{\omega',{\bm k}_\perp'}^{\rm R \dagger}|0,{\rm M}\rangle
\nonumber\\
&&~~={e^{\pi\omega/a}\over e^{2\pi\omega/a}-1}\delta(\omega-\omega')\delta^2(\bm k_\perp-\bm k_\perp') .
\label{aaac}
\end{eqnarray}
These relations reflect the identification of the left-moving modes $v_{\omega,\bm k_\perp}^{\rm F,s}(x)$
in the F-region and the modes $v_{\omega,{\bm k}_\perp}^{\rm R}(x)$  in the R-region,
and can be understood either as the entangled 
correlation between the L and R Rindler modes in (\ref{Minkowskivac})
 or as the same kind of the correlation between left-moving and right-moving modes in the F-region in (\ref{MIII}).
Using Eq.~(\ref{aaac}), we have
\begin{eqnarray}
  &&\hspace{-3mm}\langle0,{\rm M}|\phi_\F^\II(x)\phi_\R(z(\tau))|0,{\rm M}\rangle
  =\int_0^\infty d\omega\int d^2k_\perp\Bigl(v^{\F,\II}_{\omega\bm k_\perp}(x)  
    \nonumber\\
  &&
    v^{\R}_{\omega\bm k_\perp}(z(\tau))+v^{\F,\II*}_{\omega\bm k_\perp}(x)v^{\R*}_{\omega\bm k_\perp}(z(\tau))    \Bigr){e^{\pi\omega/a}\over e^{2\pi\omega/a}-1}  .
\label{vvcca}
\end{eqnarray}
Then we can
show that $\langle \phi_\F^{\rm d}(x)\phi_\R(z(\tau)) \rangle$ is reduced to the first term in parentheses in Eq.~(\ref{resultoffs}).

Remembering that the field $\phi_\F^{\rm d}(x)$  comprises the right-moving modes in the F-region coming from the L-region, 
we can definitely say that the 
{\it entanglement-induced quantum radiation} is responsible for the entanglement between the modes in the L-region
and those in the R-region or the entanglement between the left-moving and right-moving modes in the F-region. \\

{\it Summary.}---
Now it is clear why the existence of a net radiation flux from an Unruh-de Witt detector 
is in agreement with our intuition that
flux should be canceled out in an equilibrium state in a thermal bath.
The Unruh-de Witt detector is described as a thermal system only when we can integrate out the modes in the L Rindler wedge;
a typical example is the Unruh effect, the thermal behavior observed by a uniformly accelerated object in the R-region.
In the F-region, however,  this is not the case. 
The quantum field $\phi(x)$ in the F-region contains both the modes coming from the L and R regions.
If we neglect $\phi_\F^{\rm d}(x)$, the rest of the modes come from the R-region,
and we can safely integrate these modes in the L-region. 
But in order to calculate a two-point correlation function in the F-region, we cannot neglect such right-moving modes, $\phi_\F^{\rm d}(x)$, in the F-region; indeed, {\it entanglement-induced quantum radiation} is generated owing to the entanglement of $\phi_\F^{\rm d}(x)$ and the field $\phi_\R(z(\tau))$ 
on the trajectory in the R-region. 
The quantum radiation produced by a charged particle undergoing uniformly accelerated motion 
 \cite{OYZ15,OYZ16}  also has the same origin.   
Therefore, detection of the non-vanishing quantum radiation will be a test
of the entanglement of the quantum field.
In the case of a uniformly accelerated charged particle in which the classical Larmor radiation is dominant, 
the difference in the spectrum and the angular distribution \cite{OYZ16,IOTYZ} 
may make it possible to separate the quantum contribution from the classical one. 
A detection of such radiation will clarify nature of the entangled quantum vacuum.
 \\

{\it Acknowledgments.}---
This work is supported by MEXT/JSPS KAKENHI Grant Number 15H05895 (KY), 
23540329 and 16H06490 (SI). 
We thank J. Soda, Y. Nambu, M. Hotta, F.-L. Lin, R. Schutzhold, and W. G. Unruh,
for useful comments. K.Y. thanks A. Higuchi for crucial comments and discussions
and for his hospitality during the stay at York University.

\end{document}